\documentclass[preprint2,twoside]{hwo}

\usepackage{graphicx}
\usepackage{multirow}

\usepackage{rotating}
\usepackage{array}
\usepackage{booktabs}

\newcolumntype{C}[1]{>{\centering\arraybackslash}m{#1}}

\renewcommand{\arraystretch}{1.5}

\usepackage{xcolor}

\input{hwo.h}

\begin{document}

\title{\textbf{\LARGE The Habitable Worlds Observatory in Historical Context}}
\author{\textbf{\large Marc Postman$^{1}$ and Karl Stapelfeldt$^2$ }}
\affil{$^1$\small\it Space Telescope Science Institute, Baltimore, Maryland, USA; \email{postman@stsci.edu}}
\affil{$^2$\small\it Jet Propulsion Laboratory, California Institute of Technology, Pasadena, California, USA; \email{Karl.R.Stapelfeldt@jpl.nasa.gov}}

\affil{To be published in the proceedings:} 
\affil{``Towards the Habitable Worlds Observatory: Visionary Science and Transformational Technology."}

\begin{abstract}

We summarize the past four decades of astrophysics and exoplanet direct imaging mission concept studies, technology developments, and scientific progress that have led to the initiation of the Habitable Worlds Observatory project by NASA.  

$\phantom{}$
\end{abstract}

\vspace{2cm}
$\phantom{}$

\begin{quote}
{\it ``Somewhere, something incredible is waiting to be known."} -- Carl Sagan

{\it ``I was taught that the way of progress was neither swift nor easy."} -- Marie Curie
\end{quote}

\section{Introduction}

The Habitable Worlds Observatory (HWO) is humanity's most ambitious 
scientific endeavor to explore nearby sun-like stars for the presence of habitable rocky exoplanets and the possible presence of life.  In designing HWO to address this age-old question, we will also be creating a space telescope that will revolutionize our understanding of a broad range of astrophysics.  
The science cases and technical discussions in these proceedings document some of the key driving ideas that will inform the design and performance of HWO. Equally important, these articles represent input from a very wide range of astrophysics and planetary science researchers, aerospace engineers, optical systems designers, and space agency representatives. As with previous successful great observatories, the HWO architecture and its strong and growing support by the astronomical community, the public, industry, and partner space agencies are built upon decades of effort by hundreds, if not thousands, of people. In this foreword, we summarize the key scientific discoveries, technological breakthroughs, and insightful programmatic developments from the 1980s to the present that have brought us to this remarkable point in time.  

\section{The First Great Observatories and the Birth of Observational Exoplanet Science (1980 - 2005)}

Planetary exploration has always been a highly visible component of NASA science, a focus for many in the research community, and an inspiration for the public. It was inevitable that its attention would expand to include the possibilities for studying planets beyond the solar system.  An early effort in this direction was the Project Orion study \citep{Black1980}, which considered the direct detection problem at optical and infrared wavelengths and recommended a stellar interferometer as the solution.  A major impetus for progress was the discovery of the beta Pictoris circumstellar disk \citep{ST1984}, the first image of an exoplanetary system.  This supported the interpretation of stellar infrared excess emission as arising from circumstellar disks, the identification of very young stars as the hosts of protoplanetary disks, and efforts to image more systems through ground-based surveys and technological developments.  At the Jet Propulsion Laboratory, internal investments led to the development of the Circumstellar Imaging Telescope mission concept, a space telescope with an optical coronagraph.  It was soon recognized that achieving the image contrast needed to detect exoplanets in visible reflected light (10$^{-9}$ for an exo-Jupiter, 10$^{-10}$ for an exo-Earth) would require not only occulting masks, but very smooth telescope optics to focus as much starlight as possible on those masks \citep{Terrile1989}.

Started in the late 1970s, NASA's Great Observatory program (nicely described in \cite{GreatObs1986}) offered a new way to explore the entire universe -- from the solar system to the most distant galaxies -- enabled by opening new space-based windows across the electromagnetic spectrum and where all their data would be made publicly available to the world-wide astronomical community. All four of the Great Observatory missions would lead to many fundamental breakthroughs across a broad range of astrophysics. Equally importantly, these missions would demonstrate that we could build and operate precision optical systems in space with stable pointing systems and with data distribution systems that made the observations available to anyone. But it would be the {\it Hubble Space Telescope} (HST) launched in 1990 and the {\it Spitzer Space Telescope} launched in 2003 that would have especially important impacts on observational studies of exoplanets. 

In September 1989, just seven months prior to HST's launch, a symposium was held at the Space Telescope Science Institute (STScI) entitled "The Next Generation Space Telescope" \citep{NGST1980}. This was not a reference to HST nor to the as yet unimagined {\it James Webb Space Telescope} (JWST) but rather a reference to 8m - 16m telescopes deployed either in near-Earth orbit or on the surface of the moon. In the proceedings, Roger Angel discussed the potential for a lunar-based 16m telescope to directly detect Earthlike planets within 10 pc from the Sun. It was certainly audacious for the community to contemplate such capabilities at that time - but we're glad that they did - such bold thinking is essential when envisioning world-changing science.

However, for direct imaging of exoplanets, HST's optics would not be smooth enough, even without the spherical aberration discovered after launch \citep{BB1990}. At the time, manufacturing a mirror to the required smoothness appeared infeasible, and exoplanet space mission concepts began to turn away from direct imaging and toward astrometry: the Astrometric Imaging Telescope with a modest single aperture, and then to the Space Interferometry Mission (SIM).  SIM was later endorsed to move forward by the 2000 Decadal Survey of Astronomy \& Astrophysics \citep{Astro2000}.

The discovery of the hot Jupiter 51 Peg b from radial velocity measurements \citep{51Pegb1995} gave still more impetus for establishing a programmatic vision for NASA exoplanet work.  The 1993 Towards Other Planetary Systems \citep{TOPS1993}
and 1996 Exploration of Neighboring Planetary Systems \citep{ExNPS1996} reports 
suggested broad strategies to employ a variety of methods for exoplanet detection and characterization.  These recommendations helped motivate NASA investment in the Keck Observatory, and expanded research support for radial velocity work. Direct imaging was not dead, however: Taking advantage of fact that the contrast to detect an exo-Earth was 1000x less challenging in the thermal infrared, \cite{AW1997} proposed a nulling interferometer mission that later became known as the Terrestrial Planet Finder (TPF).  The 2000 Decadal Survey of Astronomy \& Astrophysics endorsed technology development for TPF as a major initiative.

New developments quickly changed the landscape for TPF.  The advent of adaptive optics technology meant that a telescope primary mirror did not need to be polished to penultimate smoothness, the effects of residual surface errors on the wavefront could be corrected by analogy to the way that Hubble's spherical aberration was corrected.  \cite{Malbet1995} showed in principle how a deformable mirror (DM) could be used to clear a dark zone adjacent to the stellar image, within which faint exoplanets could be detected in visible light.  This concept formed the basis of a 1997 proposal for a new Hubble instrument CODEX (PI R.A. Brown), which was not selected.  In the wake of CODEX, the community began a series of proposals for modest $\sim$ 1.5m class space telescopes to image exo-Jupiters using coronagraphs and DMs.  The 2000 Eclipse Discovery mission proposal \citep{Trauger2003} was not selected either, but was awarded technology development funding that led to the founding of JPL's High Contrast Imaging Testbed (HCIT) that has supported coronagraphy technology demonstrations ever since.  

When NASA funded four TPF mission architecture studies in 2001, a coronagraph study was included \citep{BATC2002} and judged sufficiently promising that the overall TPF effort was split into two tracks: the original interferometer idea (now TPF-I), and the new coronagraph option (TPF-C).  From 2002-2007 both studies proceeded under the umbrella of NASA's Navigator Program, each producing mission study reports and defining the technology development paths needed.  TPF-C was conceived as an unobscured telescope with an 8$\times$3.5 m elliptical monolith primary mirror (see Figure~\ref{fig:tel_concepts}), a concession to the limitations of launch vehicle fairings of the day.  The secondary mirror would have been deployed, and a large V-groove sunshade would have been used to provide thermal stability.  Parallel coronagraph channels using band-limited Lyot masks would have been used. There was also a general astrophysics camera that would primarily operate in parallel with long stares on exoplanet targets.  A new NASA administrator and reallocation of NASA science funds to the Ares V heavy lift launch vehicle program along with limited community support led to both TPFs, and the SIM mission, being halted in 2007.  The TPF-C final study report \citep{TPFC2006} anticipated many of the goals and technology issues HWO faces today.

In 2009 NASA established the Strategic Astrophysics Technology (SAT) competed grant program and its focus area known as Technology Development for Exoplanet Missions (TDEM), which enabled investigators to bring new coronagraph instrument ideas to the HCIT at JPL.  An important laboratory milestone was the \cite{TT2007} demonstration of $6\times10^{-10}$ image contrast a few beamwidths away from a simulated star, using a band-limited Lyot coronagraph in narrowband light.  

\section{Expanding Support: Joint Astrophysics -- Exoplanet Imaging Mission Concepts (2006 - 2015)}

As innovative as TPF-C was, it did have a major weakness: its science drivers did not appeal to a broad community of researchers. This was in large part because its design was very focused solely on the challenge of directly imaging an exoEarth. Its primary mirror did not have UV sensitivity (primarily over concerns about coating uniformity and polarization issues) and the highly elliptical PSF of the 8-m class TPF-C would be a major challenge for most general astrophysics observations. But TPF-C had the killer app -- the search for life beyond the solar system. In the ten years starting in 2006, a number of initiatives began that would address the challenge of designing a space telescope that could accomplish {\it both} the characterization of the atmospheres of Earth analogs around nearby solar type stars {\it and} perform a range of pioneering general astrophysics programs that require sensitivity from 100 nm to at least 2 $\mu$m. 

\cite{green2006} published a science case white paper for a 10-meter class UVOIR space telescope, which was a revisit of the Modern Univserse Space Telescope concept first explored by \cite{Shull1999}. One of the motivations for the \cite{green2006} article was the concern that HST would likely not last another 5 years, and thus the time was right to begin considering successors. A year later, leaders within the NASA Astrophysics Division recognized that with the 2010 Decadal review just 3 years away it would be important to support community investigations into mission concepts and technology development that the National Academies Decadal Review could potentially recommend for the coming decade. NASA thus initiated a call for Astrophysics Strategic Mission Concept Studies (ASMCS). Five studies were funded starting in 2008 that would have direct relevance to HWO: one large mission concept technology development plan: the Advanced Technology Large-Aperture Space Telescope (ATLAST), and 4 medium class mission concepts for coronagraphic direct imaging facilities: ACCESS, PECO, EPIC, and DAViNCI. 

\begin{figure*}[t!]
\centering
\includegraphics[keepaspectratio,width=6.5in]{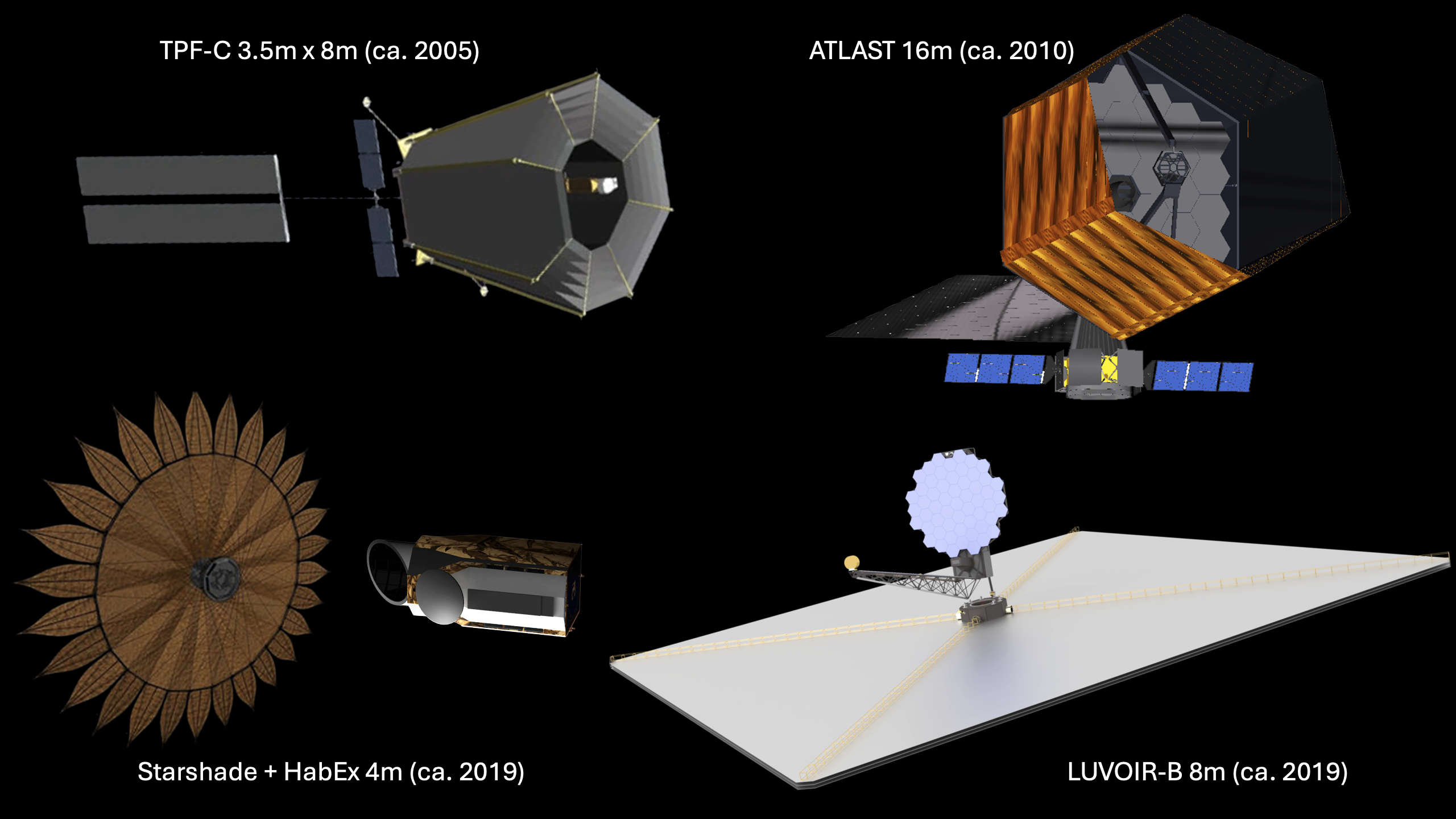}
\caption{Concept drawings of four of the large-aperture direct imaging space telescopes envisioned and studied over the past 25 years. Note that the separation depicted here between the starshade and the HabEx spacecraft is not to scale. The starshade would typically be tens of thousands of km from the telescope.}
\label{fig:tel_concepts}
\end{figure*}

The ATLAST study \citep{ATLAST2009, ATLAST2010, Feinberg2014} built upon the excitement in the exoplanet community for a flagship-class direct imaging mission but raised the astrophysics objectives of the mission to equal footing. ATLAST explored three architectures -- an 8-m monolithic space telescope and two segmented mirror concepts -- a 9.2-m telescope and a 16-m telescope (see Figure~\ref{fig:tel_concepts}). The 8-m monolithic mirror telescope could support an off-axis design. The segmented concepts were on-axis telescopes. All the ATLAST concepts had UV sensitivity down to 100 nm and were envisioned as non-cryogenic telescopes with on-orbit servicing capability. 
The ATLAST concept study identified 9 areas for technology development including UV detectors, high performance coronagraphs, precision segment actuation, wavefront sensing capable of figure knowledge to less than 5 nm rms, lightweight mirror segment manufacturing, and starshade occulters. The overall budget that the study recommended to the Astro 2010 Decadal for the ATLAST technology development plan was in the range \$287M $-$ \$335M over 10 years. The ultimate vision was that if the key technologies could be developed to TRL 6 by 2020, the community could be ready to recommend an ATLAST-scale flight mission to the Astro 2020 Decadal.

The ATLAST study was complemented by the 4 medium class space-based coronagraphic instrument development projects funded as part of NASA's ASMCS. The Pupil Mapping Exoplanet Coronagraphic Observer (PECO; \cite{PECO2010}) explored the requirements for phase induced amplitude apodization (PIAA) on a 1.4-m off-axis telescope to achieve $10^{-10}$ contrast at $2\lambda/D$ inner working angle (IWA). The Extrasolar Planetary Imaging Coronagraph (EPIC; \cite{EPIC2010}) explored the requirements for a visible nulling coronagraph (VNC) on a 1.65-m telescope to achieve $\leq 10^{-9}$ contrast at IWA $\leq 2\lambda/D$. The Dilute Aperture Visible Nulling Coronagraph Imager (DAViNCI; \cite{DAVINCI2010}) also explored a VNC instrument but envisioned its use in conjunction with four 1.1-m class mirrors on the same spacecraft to achieve high sensitivity.  All three studies targeted a small inner working angle, which is especially important since it maximizes the discovery space for imaging exoplanets (for a given telescope aperture size), while also posing greater challenges for telescope stability.  

The Actively Corrected Coronagraph for Exoplanet System Studies (ACCESS; \cite{ACCESS2008}) focused on evaluating a range of coronagraph designs (including Lyot, shaped pupil, and PIAA) with respect to science capability (e.g., contrast at the IWA, throughput efficiency, and spectral bandwidth), engineering readiness (including maturity of technology, instrument complexity, and sensitivity to wavefront errors), and mission cost. 

In retrospect, NASA's funding of many different medium-class coronagraph mission concepts with similar telescopes but different starlight suppression approaches resulted in duplication of effort and fragmentation of the community.  Exploration of coronagraph types at the subsystem level, not the mission level, would have been more appropriate.  A proper trade study of coronagraph types would have to wait for the Astrophysics Focused Telescope Asset (AFTA) Coronagraph Working Group activity, and the Exo-C study \cite{ExoC2015}.  There was also programmatic disagreement on whether the science return of a medium-class mission would be sufficient to justify the $\sim$\$1B cost, and whether a medium class mission would distract from the larger goal of reaching for exo-Earth imaging on a larger-scale mission.  

Nearly contemporaneously with the NASA ASMCS studies, the NASA Institute for Advanced Concepts (NIAC) funded an initiative to investigate the viability of an external occulter system or starshade \citep{Cash2006}. A mission concept based solely upon the use of a starshade, called the New Worlds Observer, was performed \citep{NWO2007, Cash2008} and was submitted for consideration to the Astro2010 decadal review.  The use of a starshade avoids the strict telescope stability requirements that come with the use of coronagraphy to achieve very high contrast, but adds the complexity of a precision deployed structure and a second spacecraft. 

While the above studies explored coronagraph or starshade viability for medium class missions, they all have relevance to large aperture telescope concepts like those studied in ATLAST as well as to the current designs for HWO. 

There were several scientific milestones in exoplanet research in the first decade of the 21st century that helped pave the way for HWO. The STIS spectrograph on HST was used to acquire the first spectrum of an exoplanet atmosphere \citep{Charbonneau2002}. The first direct imaging of exoplanets using a ground-based adaptive optics system was accomplished \citep{Marois2008}. And in 2009, the {\it Kepler} mission \citep{Kepler2010} was launched. The {\it Kepler} space telescope was NASA's first space-based mission dedicated to the search for exoplanets via the transit method. It would ultimately lead to the discovery of over 5000 exoplanet candidates and place new constraints on the frequency of planets around stars, in general, and of $\eta_{Earth}$, in particular. 

In the area of general astrophysics, the installation of the Cosmic Origins Spectrograph (COS; \cite{COS2012}) on HST in 2009 opened up an era of high sensitivity UV spectroscopic studies of stars, star-forming regions, and the circumgalactic medium surrounding galaxies. Many of the resulting COS programs achieved remarkable improvements in characterizing interstellar chemical enrichment, star formation physics, and galaxy evolution that were uniquely enabled with a high performance UV spectrograph.

The 2010 Astronomy and Astrophysics Decadal Survey report showed the influence of the scientific and technical progress during the preceding decade by its choice of title "New Worlds, New Horizons" \citep{Astro2010}. The recommendations of Astro2010 gave top priority in its medium scale category to a \$200M technology development program for a future optical space telescope exoEarth direct imaging mission, leaving the interferometer concept by the wayside. The Astro2010 committee also gave top priority in its large space mission category to the Wide-Field Infrared Space Telescope (WFIRST), which ultimately became the {\it Nancy Grace Roman Space Telescope}. However, the WFIRST mission as conceived of in Astro2010 did not include a coronagraph and was initially proposed as a 1.5-m telescope. The small WFIRST aperture was driven in part by the requirements from several probe scale dark energy mission studies performed earlier that decade \citep{SNAP2004, Destiny2005, JDEM2010} and from funding constraints imposed by NASA.

Early in 2011, the National Reconnaissance Office (NRO) informed NASA that two 2.4-m optical telescope assemblies were available for NASA's use. This offer enabled NASA to potentially implement a larger version of WFIRST, which was compelling. Firstly, it would allow WFIRST to achieve greater angular resolution and depth than the recently approved ESA Euclid mission to launch a 1.2-m wide-field telescope to survey the sky from space. Both missions could be complementary but at its original 1.5-m aperture size, WFIRST ran the risk of being redundant. Secondly, it potentially offered NASA some cost savings as they would not need to build WFIRST from scratch. Lastly, it provided a nice opportunity for NASA to demonstrate it could make good use of a powerful but no longer needed defense asset. NASA accepted the offer in the summer of 2011. In 2012, the members of the WFIRST Science Definition Team (SDT) wrote a white paper \citep{SDT2012} intended for NASA HQ that highlighted the scientific advantages of using the NRO-2.4m for WFIRST. In \cite{SDT2012} the SDT proposed that a coronagraphic imager could be included on WFIRST {\it "for direct detection of dust disks and planets around neighboring stars, a high-priority science and technology precursor for future ambitious programs to image Earth-like planets around neighboring stars."}  Adding a coronagraph instrument to an already approved space telescope would be much cheaper than flying a separate mid-scale mission focused only on coronagraphy. The community would get the mission it had been long proposing, albeit with a sub-optimal telescope and only as a technology demonstration with no science requirements.

Although the 2010 decadal review recommended significant technology development funding for a future exoEarth imaging space telescope, the actual investment dispensed by NASA over the 2010-2019 decade for {\it generic} direct imaging technology and ultra stable wavefront sensing and control systems was less than the \$200M suggested (the amount was closer to \$100M). However, once John Grunsfeld, Associate Administrator of NASA's Science Mission Directorate at the time, approved the inclusion of a WFIRST coronagraph in 2014, a significant investment ($\sim$\$300M) focused specifically on that instrument was made. 
This highlights a key lesson that NASA seems most effective at investing in technology development for mission-specific needs rather than generic visions. 
We refer the reader to the excellent article within these proceedings by Pueyo, Macintosh, \& Laginja on how the coronagraphic instrument (CGI) on the {\it Roman Space Telescope} proved to be a significant technology pathfinder for HWO coronagraph technology.

The HCIT at JPL has been focused on testing coronagraphs that are used in conjunction with monolithic mirror telescopes and HCIT played a key role in testing coronagraph concepts for the {\it Nancy Grace Roman Space Telescope}. An independent testbed focused on PIAA coronagraph technology development had also been established early in the new millennium \citep{piaa2005, guyon2010}. In 2013, the Russell B. Makidon Optics Laboratory (RMOL) was established at the STScI with a focus on testing coronagraphs that could work with complex aperture telescopes, with a specific focus on segmented apertures. The RMOL testbed for coronagraphs was called the high-contrast imager for complex aperture telescopes (HiCAT; \cite{RMOL2013}). The RMOL and HiCAT were direct outcomes of the more generic investments that NASA made in coronagraph technology development. Both HCIT and HiCAT have played, and will continue to play, critical roles in validating the contrast achievable using the current state-of-the-art in coronagraph designs compatible with HWO. 

In 2014, the Association of Universities for Research in Astronomy (AURA) sponsored a year-long study of a future UVOIR space telescope. This concept, called the High-Definition Space Telescope (HDST), was a 12-m segmented aperture telescope. As with ATLAST, HDST would be capable of both direct imaging of a significant sample of exoEarths as well as performing a broad range of general astrophysics observations. While the AURA study did not have the resources to perform detailed engineering design work for HDST it did significantly expand the general astrophysics science use cases, which had a sizeable influence on the subsequent LUVOIR and HabEx investigations later in the decade. The HDST report \citep{HDST2015} helped keep the community momentum for a large UVOIR mission going in the mid-decade timeframe. 

At the same time as the HDST study was underway, two highly influential investigations of the exoEarth yields expected for direct imaging missions were published \citep{Stark2014, Stark2015}. These demonstrated that the exoEarth yield depends on a range of instrumental, astrophysical, and observational parameters. One of the strongest dependencies was on telescope aperture, D, with the exoEarth yield scaling as $\rm D^{\sim 2}$. The Stark et al. studies were influenced by work led by Robert Brown \citep{Brown2004, Brown2010} who explored the impact of observational completeness on exoEarth imaging efficiency but the Stark et al. studies went into more depth and developed an optimized star selection algorithm to maximize yield. Furthermore, these yield results were in contrast to an earlier work by \cite{Beckwith2008} who concluded that yield scaled as $\rm D^3$ but only if a volume-limited survey of habitable zones in nearby stars could be performed. The feasibility of performing a volume-limited survey proposed by Beckwith did not take into account observational completeness and the various factors addressed in the more thorough analysis done by Stark et al. It thus became very clear that the aperture size of a future direct imaging telescope would be a very important factor in determining the likely success of a mission in the search for life beyond the solar system given that the $\rm D^2$ scaling of exoEarth yield was the realistic expectation. 

As important, the \cite{Stark2014} study provided a framework for assessing the performance of direct imaging telescope designs in an objective way and also inspired the development of a range of yield estimation algorithms \citep{Yield2014A, EXOSIMS2016, Yield2018A, Yield2019A, Yield2019B, Yield2020A, Morgan2021}. All subsequent direct imaging concepts, including HDST and the large direct imaging missions studied in the pre-2020 decadal mission studies (HabEx, LUVOIR), used one or more of these yield calculations to test and compare their performance in detecting and characterizing potentially habitable worlds. 

\section{Realizing the Mission (2016 - 2025)}

In January 2016, NASA's Astrophysics Division initiated a call to the community to seek participants on Science and Technology Definition Teams (STDTs) for the Astro2020 Decadal Candidate Missions. Specifically, four mission concepts would be studied with two specifically focused on the science cases and architectures for a combined exoEarth imaging and general astrophysics observatory operating over the UV through NIR wavelength range. Those two were the Habitable-Exoplanet Imaging Mission (HabEx) and the Large Ultraviolet, Optical, and Infrared Surveyor (LUVOIR). While these initiatives were similar in intent as the 2008 ASMCS, the 2016 initiative was significantly better funded and the experience of the past decade had given the STDTs a significant head start. Team chairs were selected by early spring, followed by selection of the team members. The teams began their work in fall of 2016.

While the HabEx and LUVOIR teams had significant scientific overlap their charge from NASA's Astrophysics Division was different. The HabEx STDT was tasked with identifying a mission that could fit within a \$5B cost cap. The LUVOIR STDT was asked to be fully science driven within the realm of viable technology and explore telescope concepts with apertures of at least 8-m. 
A common challenge recognized by both teams based upon the work done since TPF-C was that the performance of the telescope had to be considered at the system level. One cannot achieve contrast levels of $10^{-10}$ by just designing a coronagraph and attaching to a telescope as an independent instrument. The entire observatory {\it is} the direct imaging system - from the stability and smoothness of the optical surfaces, to the choice in secondary mirror size and placement, to the active wavefront sensing and control and thermal stability systems, to the uniformity of the UV coatings, to the disturbance isolation between the telescope and spacecraft, and, of course, the internal optics and design of the coronagraph and/or performance of the external starshade.  The telescope architecture must be subordinate to the requirements of high contrast imaging, not the other way around.  Both teams went in with eyes wide open to these challenges.

The HabEx STDT explored 9 architectures ranging in aperture size from 2.4m to 4m. However, they found that the exoplanet yield for telescopes with apertures less than 4m were too low (less than 3 exoEarths) to achieve all of their top scientific objectives. Their final preferred recommendation was thus the 4m design (Figure~\ref{fig:tel_concepts}), which used both an internal coronagraph and an external starshde \citep{HabexFinal2019}. 

The LUVOIR STDT found that even with a higher aperture range, the telescope design was critical to ensuring high exoEarth yields. The initial LUVOIR-A 15m on-axis design with a secondary mirror that had an obscuration of the primary pupil in excess of 10\% resulted in a significant reduction in exoEarth yields that negated the aperture advantage. The final LUVOIR-A on-axis design \citep{LUVOIR2019} had a central obscuration of just 1\%, resulting in much higher yields. LUVOIR-B was an 8m off-axis system and thus had no central obscuration. It is worth noting that the main limitation to the aperture size on the off-axis LUVOIR-B concept (Figure~\ref{fig:tel_concepts}) were the likely launch vehicle and fairing combinations that could accommodate it.

The funding for the STDTs included significant resources for cost estimates to be performed at a level of detail unprecedented for previous pre-decadal mission studies. This was a real programmatic advantage of these studies, particularly in light of the cost overruns associated with JWST -- which were clearly on the minds of both the decadal review committee members and NASA. The STDT cost estimates ultimately were in very good agreement with the estimates based on the independent cost models employed by the Decadal review. This was a validation that given sufficient technical detail, one can get far more realistic costs for flagship scale missions. 

In-space servicing was extensively explored by the LUVOIR STDT. Modular instrument and avionics bays with easy external access points potentially support multiple generations of scientific instruments and ensure a multi-decade lifetime for the observatory.  Servicing provides more flexibility in developing instrument capabilities -- features that may prove too daunting for the first generation of instruments (e.g., a UV corornagraph, a UV polarimeter) could be implemented as second generation instruments 3-5 years after launch. The modularity required for servicing also has benefits for simplifying integration and testing prior to launch. 

One of the most positive aspects of the composition of the HabEx and LUVOIR STDTs was the close collaboration between the two teams. This led to sharing of ideas and keeping each team updated on progress. Even though there was a natural competition between the teams, this good working relationship was important. An especially important unwritten rule was that neither team would denigrate the other mission concept. This ultimately helped both teams to present a unified vision to the decadal review committee that there was strong support, regardless of the mission specifics, within the community for an observatory capable of direct imaging of Earth-like exoplanets that would also enable ground-breaking general astrophysics studies.

In addition to the HabEx and LUVOIR reports, progress in the HCIT was also an important input to the Astro2020 Decadal Survey.  \cite{Seo2019} showed a laboratory demonstration of 4$\times10^{-10}$ contrast in 10\% bandwidth at 3$\lambda/D$ separation from a simulated star, in a 360$^{\circ}$ dark hole using a classical Lyot coronagraph and an unobscured monolith telescope pupil.  Although the dark hole was smaller than HabEx or LUVOIR requirements, this result proved crucial for showing the Astro2020 committee that the needed starlight suppression performance was approaching tbe requirements of an exo-Earth direct imaging mission. 

Theoretical progress was made by the RMOL HiCAT team with the introduction of a new class of Apodized Pupil Lyot Coronagraphs (APLC) capable of working with segmented aperture telescopes to remove broadband diffracted light from a star with a contrast level of $10^{-10}$ \citep{APLC2016}. This development helped grow confidence that large aperture segmented telescopes had the potential to be used for direct detection of ExoEarths around solar type stars. In 2024, with funding from NASA's SAT program, the HiCAT team used an actual apodized Lyot coronagraph to achieve a contrast of $6 \times 10^{-8}$ with a segmented aperture in a 9\% bandpass in a 360 deg dark hole with inner and outer working angles of 4.4$\lambda/D$ and 11$\lambda/D$ \citep{Soummer2024}. HiCAT also achieved a narrowband contrast (3\% bandpass) of $2.4 \times 10^{-8}$. These performance levels were achieved with the testbed in an ambient air environment. While these results are clearly not yet at the level needed for exoEarth imaging, this progress is significant given that even as recently as 2010 a number of experts in high contrast imaging did not think such performance was possible with a segmented aperture system. This demonstration builds upon a number of recent innovations developed at the RMOL \citep{RMOL2020, Por2020, Pueyo2022, Laginja2022, RMOL2023a, RMOL2023b, RMOL2024}.

On the scientific front, the {\it Spitzer Space Telescope} verified the existence of 2 of the known planets in the TRAPPIST-1 system and discovered 5 more \citep{Gillon2017}, making this system the one with the largest number of known planets in its habitable zone. And in 2018, the Transiting Exoplanet Survey Satellite (TESS) was launched. TESS \citep{TESS2015} was designed to survey the brightest stars all around the sky for transiting exoplanets, targets that would provide the best signal levels for spectroscopic study of their atmospheres with JWST.  

Prior to Astro2020, an NAS initiative entitled the ``Exoplanet Science Strategy" was undertaken to layout guidance for key initiatives in the field of exoplanet research \citep{ESS2018}. A key recommendation of the report was that {\it ``NASA should lead a large strategic direct imaging mission capable of measuring the reflected-light spectra of temperate terrestrial planets orbiting Sun-like stars."}

The Astro2020 decadal review outcomes were released in early November 2021 \citep{Astro2020}. The NAS recommended what is now referred to as HWO as NASA's top priority for its large mission category. Specifically, the report stated {\it ``The decadal survey recommends a large ($\sim$6m diameter) Infrared/Optical/Ultraviolet space telescope with high-contrast imaging and spectroscopy as the first mission to enter the Great Observatories Mission and Technology Maturation Program. This is an ambitious mission with the goal of searching for biosignatures from habitable zone exoplanets and providing a powerful new facility for general astrophysics. If mission and technology maturation are successful, as determined by an independent review, implementation should start in the latter part of the decade with a target launch in the first half of the 2040’s."}

Just 7 weeks after the release of the Astro2020 report, on Christmas Day 2021, the global astronomy community along with our colleagues in industry, NASA, ESA, and CSA, watched breathlessly as JWST lifted off onboard an Ariane V launch vehicle bound for its Sun-Earth L2 halo orbit centered 1.6 million km from Earth. The successful launch and  deployment of JWST represented a key scientific and technological step towards HWO. A large 6.5-m deployable segmented space telescope was now at TRL9 albeit one with 60 nm rms wavefront stability instead of the ten picometer stability that will be needed for HWO. Nonetheless, JWST's on orbit performance in angular resolution and stability exceeded its requirements by about a factor of 2. The era of large deployable precision space telescopes was here. And so was an era of remarkably new exoplanet science. JWST was not just a boon for understanding the early universe. Its ability to perform transit spectroscopy with unprecedented sensitivity and spectral resolution of both gas giants and rocky worlds around M-dwarf stars has revealed many new insights about the abundance of water, CO$_2$, CH$_4$ and other species in exoplanet atmospheres. JWST also carried the first space-based near-IR IFU and first space-based multi-shutter assembly, instrument types that may be implemented for HWO but with peak performance in the UV and optical wavelength regimes.

As noted in the Astro2020 report, a technology maturation program was a top recommendation in addition to the actual direct-imaging mission. NASA adopted a rather slow response following the 2020 decadal report's release to initiate such a program. The Science, Technology, Architecture Review Team (START) and the Technical Assessment Group (TAG) were the initial forms this response took. While the level of funding was far below what a vigorous technology maturation program would need, the START and TAG did do an excellent job of rallying the community and a number of the key technical designs and science use cases in these proceedings were initially begun as part of the START and TAG effort. 

In 2023, at a ceremony commemorating the recent 50th anniversary of the last Apollo mission to the moon, NASA Administrator Bill Nelson named the Astro2020 Decadal UVOIR space telescope mission the ``Habitable Worlds Observatory."  The official NASA HWO Project Office was established a year later in August of 2024.

\section{Closing Thoughts}

Over the past four decades a number of realities regarding large astrophysics missions are clear - science is a very necessary but not sufficient ingredient for mission success.  Broad community support, technical feasibility, cost, accountability, and allies in government are all essential as well. International partnerships are very beneficial and often necessary. HWO is now on a pathway to assemble all of these ingredients. New launch vehicles should provide new opportunities for realizing a robust and feasible architecture for HWO.  Newer coronagraph designs (e.g., the vortex coronagraph) and progress in picometer wavefront sensing \& control provide momentum toward solving the still very challenging objective of exo-Earth imaging in the optical band at $10^{-10}$ contrast, 20\% bandwidth, and with a segmented primary mirror telescope.  On-orbit servicing can simplify observatory I\&T and extend the mission's scientific impact for multiple generations. As of the end of 2025, the most significant influx of funds to HWO technology development has been indirect -- centered on the CGI developed for the {\it Nancy Grace Roman Space Telescope}. Very significant funding for a HWO-focused technology maturation program will be needed over the next few years to meet the demands of this mission.
But a real advantage that HWO has is that we now know what the requirements are to successfully image Earth-like exoplanets. When HST was built we did not know any of those requirements. When JWST was built we mostly knew the requirements but the mission focus was elsewhere. With {\it Roman} the use of any high contrast imaging requirements to dictate the choice of the optical telescope assembly was not an option. HWO has the chance to finally get this right, and if it does it will be a history-making scientific endeavor that truly helps us tell the story of life in the universe.

\bibliography{author}

\end{document}